\documentstyle[12pt]{article}

\begin{document}
\rightline{}
\rightline{IPNO/TH 94-64}
\rightline{(Revised version) November, 1994}
\vspace{3mm}
\centerline{\bf CHIRAL SUM-RULES FOR ${\cal L}^{WZ}_{(6)}$ PARAMETERS }
\centerline{\bf AND APPLICATION TO $\pi^0,\eta,\eta'$ DECAYS}
\bigskip
\centerline{B. Moussallam}
\smallskip
\centerline{\sl Division de Physique
Th\'eorique\protect\footnote{Unit\'e de
Recherche des Universit\'es Paris 11 et Paris 6 associ\'ee au CNRS}
, Institut de Physique Nucl\'eaire}
\centerline{\sl Universit\'e Paris-Sud, 91406, Orsay, France}
\vfill
\centerline{\bf ABSTRACT}

The chiral expansion of the low energy processes
$\pi^0\to\gamma\gamma$ and $\eta\to\gamma\gamma$ is
reconsidered with particular emphasis on the question of
the evaluation of the two low-energy parameters from ${\cal L}^{WZ}_{(6)}$
which are involved at chiral order six. It is shown how
extensive use of
sum-rules and saturation with resonances as well as constraints
from asymptotic QCD effectively determine their
values. Predictions for the widths are presented
for both standard and non-standard values of the quark
mass ratio $m_s/{\hat m}$. A
precise relation is established between the usual phenomenological
$\eta-\eta'$ mixing parameters and those of the chiral expansion.
The large size of the chiral correction to the $\eta$ decay
can be understood on the basis of a simple counting rule:
$O(1/N_c)\sim\ O(m_q)$. It is shown how this counting rule eventually
allows one to include the $\eta'$ into the effective lagrangian
in a consistent and systematic way.

\vfill
\eject
\rightline{}
\centerline{\bf 1. INTRODUCTION}

The $\eta$ electromagnetic decay is a neat low-energy process,
free of any strong final-state interaction, so one would expect
that the chiral expansion should converge as fast
as it seems to do in the case of the masses.
Yet, at leading order, the width is predicted to be too small
by a factor of three compared to experiment.
The problem of the $O(p^6)$ corrections to
$\eta\to2\gamma$ (and $\pi^0\to2\gamma$) has been addressed
several times in the
literature\cite{dhl}\cite{bbc1}(who computed the
chiral loop contribution) \cite{r+f}\cite{dw}.
Strictly speaking, chiral perturbation theory ($\chi$pt)
is unable to make any prediction in this case
because there are precisely two low-energy (LE) parameters,
which are finite and called $t_1$ and $t'_1$ below,
which cannot be determined  at present
from other low-energy data (in principle the amplitudes
for $\eta\to\pi^0\pi^0\gamma\gamma$ or $\gamma\gamma\to
3\pi^0$ could be used as they contain just the same parameters).
However, shortly after the
$\chi$pt has started to be developed in a systematic way
\cite{gl84}\cite{gl85} along with its list of LE parameters,
a great deal of progress was made in the art of relating,
in a rather precise way, the size of these parameters to the
properties of the low-lying resonances\cite{egpr}\cite{eglpr}.
In particular, as \cite{eglpr} have  emphasized, whenever one
can write down a rapidly converging dispersion relation
it is natural to expect that saturation from the first
low-lying resonances should provide an adequate approximation.

In this work, we intend to exploit in this manner specific
convergence properties of dispersion relations, which
can be shown to hold in QCD in the chiral limit, in order
to estimate the two LE parameters which occur in the
electromagnetic decay of the $\pi^0$ and the $\eta$. These
sum-rules are similar to the ones that were established for
$L_{10}$\cite{l10} and for $L_7$\cite{gl90} on the basis
of two-point functions. The parameters $t_1$ and $t'_1$ are
related to three-point functions instead. We will find out that
$t'_1$ is proportional to the square root of $-L_7$
and to the amplitude
$A({\eta'}\to\gamma\gamma)$. This amplitude must be taken from
experiment and this information feeds back into the LE
parameter, as is usual for
the properties of the light resonances\cite{egpr}.
Similarly, the parameter $t_1$ may be shown to encode
information on the $\pi(1300)$ resonance electromagnetic
decay\cite{r+f}. Fortunately, (because hardly anything is
experimentally known on this decay channel) we show that
this contribution
is numerically dominated by a contribution from the asymptotic
behaviour of the three-point function which turns out to be
canonical. A similar behaviour was exploited previously in the
case of three-point functions involving one scalar current and
was claimed to provide an explanation, based on chiral symmetry,
for the somewhat unexpected electromagnetic properties of the
scalar mesons\cite{ms}.

Besides the purely phenomenological application,
once it is ensured that the LE parameters
are estimated to a reasonable level of accuracy, one can address
the  question of the rate of convergence
of the expansion in powers of the quark masses. The conventional
determination of the light quark running masses $m_u$ and $m_d$
\cite{gl82}
was recently challenged\cite{ssfgt}, based on an analysis of
the violation of the Goldberger-Treiman relation.
This analysis suggests that the value of the quark mass ratio
$2m_s/(m_u+m_d)$ is two to three times smaller than in the
standard $\chi$pt, i.e. $r\simeq r_2=25.9$.
If true, this
necessitates to rearrange the quark mass terms in the chiral
perturbation series in a
different way\cite{ssfpl}\cite{ssfpr}\cite{daphne}
(the so-called generalized $\chi$pt) which one would expect to
converge more rapidly.
In  practice, distinguishing between the $\chi$pt and its
generalized variant
is, curiously, not so easy. It is likely that only very precise
measurements of $\pi-\pi$ scattering lengths could settle the issue
of which one is correct. It turns out that the amplitude
$A(\eta\to\gamma\gamma)$ is very sensitive to the value of $r$
when $r$ is in the vicinity of the value $r_2$. Nevertheless, we
will show that it is also possible to reproduce the experimental
results under the assumption of a much smaller value of $r$.

The plan of the paper is as follows. The next section contains
the derivation of the sum-rules for the parameters $t_1$ and $t'_1$,
the main results are contained in the
formulae (\ref{eq:t1t2}) and (\ref{eq:t'1}).
Application to the $\eta$ decay amplitude is then discussed in sec.3.
In particular, we establish a connection between the chiral expansion
description and the phenomenological representation in terms of an
$\eta-\eta'$ mixing angle which is widely employed (e.g. \cite{gk}).
The chiral corrections to the $\pi^0$ decay are also worked out
here.
This analysis suggests that the value of $F_{\pi^0}$ quoted in the
literature is incorrect.
The description of the
$\eta$ decay amplitude in the generalized $\chi$pt is presented at
the end of sec.3. The $\eta-\eta'$ system is of interest also in
connection with the large $N_c\ $ expansion since, in the large $N_c\ $
limit, the $\eta'$ is a ninth Goldstone boson. This suggests still
another expansion scheme where 1/$N_c\ $ is considered as an expansion
parameter together with the quark masses and the momenta. A natural
choice is to count one power of 1/$N_c\ $ on the same footing as
one power of a quark mass. This point of view is discussed in sec.4
(in particular, one observes the appearance of two different
mixing angles but only one of them is relevant to the $\eta$ decay)
and the results are compared to those of the standard
chiral expansion.

\vspace{2mm}
\centerline{\bf 2. LOW-ENERGY PARAMETERS
FROM A CHIRAL SUM-RULE METHOD}
\vspace{2mm}

The current-algebra prediction for the amplitudes describing
pseudo-scalar meson decays into two photons is contained into
the following term of the canonical Wess-Zumino
lagran\-gian\-\cite{wz}\-\cite{wzw}
\begin{equation}
{{\cal L}^{WZ}_{(4)}}={N_c\over32\pi^2F_0}\epsilon^{\mu\nu\alpha\beta}
<\phi\, v_{\mu\nu} v_{\alpha\beta} >
\label{eq:wz}
\end{equation}
where the totally antisymmetric tensor is normalized
such that $\epsilon^{0123}=1$ and $\phi=\sum\phi_i\lambda_i$
,\ i=1,8. The subscript in eq.(\ref{eq:wz}) is a reminder of the
chiral order. Part of the higher chiral corrections are
contained in chiral lagrangian terms which are also
proportional to the $\epsilon$ tensor.
At order six there are three
independent terms which are relevant for our purposes (using
relatively standard notations, see e.g. ref.\cite{egpr})
\begin{eqnarray}
{\cal L}^{WZ}_{(6)}=i\epsilon^{\mu\nu\alpha\beta}\bigg\{
&&t_1< \chi^{(-)} f^{(+)}_{\mu\nu} f^{(+)}_{\alpha\beta} >+
t'_1< \chi^{(-)}>< f^{(+)}_{\mu\nu} f^{(+)}_{\alpha\beta}>\nonumber\\
&&-it_2<d_\lambda   f^{(+)}_{\lambda\mu}\{f^{(+)}_{\alpha\beta},
u_\nu\}>+...
\bigg\}
\label{eq:lag}
\end{eqnarray}
As discussed in \cite{aa}, all the other potentially relevant
terms that one can
write down can be reduced to the above three by the use of
the so called Shouten identity and the equation of motion.
Among the constants appearing in (\ref{eq:lag})
only $t_2$ has been estimated previously
\cite{pp}\cite{bbc2}\cite{bij90}.
The parameter $t_2$ is phenomenologically interesting in that it
controls the corrections to decays like $\pi^0\to\gamma\gamma^*$
which were discussed in \cite{bbc2}. This contribution
vanishes when both photons are on-shell. In this case, the terms
$t_1$ and $t'_1$ are the {\it only}
contributions from ${\cal L}^{WZ}_{(6)}$ to
the processes
$\pi^0\to 2\gamma$ and $\eta\to2\gamma$.
These two parameters are finite,
reflecting the fact that the loop contribution is finite
and, as a matter of fact, vanishes\cite{dhl}\cite{bbc1}.
The parameter $t'_1$ is
analogous to some extent to the parameter $L_7$ of the
standard $O(p^4)$ lagrangian\cite{gl85}\cite{gl90}. It picks up a
contribution from the pole of the ${\eta'}$ and, from the point of
view of the large $N_c\ $ counting, it is of order $O(N_c^2)$ instead
of being $O(N_c)$ at most, like the other parameters. From this
point of view one expects it to play a dominant role. The second
parameter, $t_1$ is unrelated to the ${\eta'}$ resonance.

In order to obtain an estimate for $t_1$, $t'_1$ and $t_2$
let us take the chiral limit and consider the
vector-vector-pseudoscalar correlation function, i.e.
\begin{equation}
d^{abc}\epsilon_{\mu\nu\alpha\beta}\,p^\alpha q^\beta\,
\Pi_{VVP}(p^2,q^2,r^2)=
\int d^4x\, d^4y\, e^{ipx+iqy}<0\vert T\Big( j_\mu^a(x) j_\nu^b(y)
j_P^c(0)\Big)\vert 0>
\label{eq:vvp}
\end{equation}
with $r=-(p+q)$, $a,b,c=1$ to 8 and where
\begin{equation}
j^a_\mu(x)=\bar\psi(x){\lambda^a\over2}\gamma_\mu\psi(x),\quad
j^c_P(x)=i\bar\psi(x){\lambda^c\over2}\gamma_5\psi(x),\quad
\end{equation}
are the vector and the pseudo-scalar currents respectively. In
the case  of the singlet pseudo-scalar current, i.e. when the
index $c$ is set to $c=0$ in (\ref{eq:vvp}) we define a function
$\Pi^0_{VVP}$ from exactly the same formula. $\Pi^0_{VVP}$
differs from $\Pi_{VVP}$ because of the presence of the
$U(1)$ axial anomaly.

Firstly, let us perform the low-energy expansion of
$\Pi_{VVP}$.
At order $O(p^6)$
one has to include a tree-level contribution from
${{\cal L}^{WZ}_{(4)}}$, a
one-loop contribution involving one
vertex from ${{\cal L}^{WZ}_{(4)}}$
and, finally, a tree contribution
involving the parameters $t_1$, $t'_1$ and $t_2$ from
${\cal L}^{WZ}_{(6)}$ (\ref{eq:lag}).
In the minimal
subtraction scheme the scale dependence which arises from
the loop contribution is cancelled by that of the parameter
$t_2$\cite{dw},\cite{bbc2}, the first two
parameters are scale independent.  The expansion is as
follows
\begin{equation}
\Pi_{VVP}(p^2,q^2,r^2)=2B_0\left\{ {N_c\over16\pi^2 r^2} + 16t_1
+4t_2\,{p^2+q^2\over r^2} \right\}+({\rm chiral\ loop})
\label{eq:vvpdev}
\end{equation}
The first term in (\ref{eq:vvpdev}) is the pole
contribution from the canonical Wess-Zumino lagrangian.
In the pseudo-scalar singlet channel, now,
we no longer have a Goldstone boson pole contribution so that
the chiral expansion starts with a constant term:
\begin{equation}
\Pi^0_{VVP}(p^2,q^2,r^2) =2B_0\left\{ 16t_1 +48t'_1 \right\}
+O(p^2,q^2,r^2)
\label{eq:vvpdev0}
\end{equation}

For moderate values of the momenta $p,q\sim 1$ GeV it is
natural to assume that $\Pi_{VVP}$ and $\Pi^0_{VVP}$ are
dominated by the low-lying vector and pseudo-scalar resonance
poles. A very useful constraint arises in the chiral limit
from the behaviour in the asymptotic regime $p,q\to\infty$.
Indeed, a  simple calculation in QCD
shows that one has\footnote{The author
is indebted to Marc Knecht for noticing a sign error in the first
version of the manuscript}:
\begin{equation}
\lim_{p,q\to\infty} \Pi_{VVP}(p^2,q^2,r^2) =
\lim_{p,q\to\infty} \Pi^0_{VVP}(p^2,q^2,r^2)
=-B_0 F^2_0\,{p^2+q^2+r^2
\over 2 p^2 q^2 r^2} (1 + O(\alpha_s) ) + ...
\label{eq:vvpas}
\end{equation}

In other terms, under a scaling of the momenta $p\to\lambda p$,
$q\to\lambda q$, $\Pi_{VVP}$ scales as $1/\lambda^4$
if we let $\lambda$ go to infinity.
The leading term in the asymptotic behaviour has the property
of being canonical, i.e. it does
not contain powers of logarithms caused by
anomalous dimensions. This is
because the scalar condensate $<\bar\psi \psi>$
has the same anomalous
dimension as the scalar (or the pseudoscalar) current and the vector
current carries no anomalous dimension.
In deriving (\ref{eq:vvpas}) we have performed the operator product
expansion \`a la SVZ\cite{svz}, i.e. taking into account the
non-perturbative feature that the quark and gluon condensates are
non-vanishing. SVZ argue that doing this enlarges the domain
of applicability of the asymptotic behaviour down to the 1-2 GeV
domain. It is therefore natural to demand that the two domains match
smoothly. This is really useful only in the chiral limit because
the correlation function does not blow up
asymptotically
and one need not resort to tricks like the
Borel transformation which are necessary in the more general
situation.
Resonance saturation in the low to medium energy region
amounts to approximate the correlation function by a meromorphic
function having simple poles at the location of the resonance
masses. Reducing to a common denominator, one can in principle
have an arbitrary polynomial in the numerator. It is here that
the asymptotic conditions come into play and limit the degree of
the polynomial. In the present case, where the asymptotic
behaviour is canonical, the term of highest order in the polynomial
gets exactly determined. Let us for instance include
one nonet of vector resonances in the vector channel and only the
pion octet in the
pseudoscalar channel. Then, we obtain a very simple representation:
\begin{equation}
\Pi_{VVP}(p^2,q^2,r^2)={-B_0F^2_0\,(p^2+q^2+r^2) + a \over
2(p^2-{M_V}^2)(q^2-{M_V}^2) r^2 }
\label{eq:vvpres1}
\end{equation}
where ${M_V}$ is the vector meson mass in the chiral limit.
This construction bypasses the use of effective lagrangians
for resonances and yields automatically
the most general form of the amplitude containing
the right poles and obeying the appropriate asymptotic
constraints. In the case above one has
a single arbitrary parameter, $a$.
In writing down (\ref{eq:vvpres1}) we have ignored
the contribution of the $\pi(1300)$  ($\pi'$)
multiplet. If we include
the $\pi'$ pole contribution into the amplitude, we obtain
a more complicated representation,
\begin{equation}
\Pi_{VVP}(p^2,q^2,r^2)={(-B_0F^2_0\,(p^2+q^2+r^2) + a)
(r^2-{M_P}^2) + b (p^2+q^2) + c r^2
\over
2(p^2-{M_V}^2)(q^2-{M_V}^2) r^2 (r^2-{M_P}^2)}
\label{eq:vvpres2}
\end{equation}
which contains two additional parameters $b$ and $c$.
${M_P}$ is the $\pi'$ nonet mass in the chiral limit.
This representation reduces to the preceeding one
in the limit where the mass ${M_P}$ is sent to infinity. It
is useful to rewrite (\ref{eq:vvpres2}) by separating out the
various poles in $r^2$
\begin{eqnarray}
&&\Pi_{VVP}(p^2,q^2,r^2)={1\over2(p^2-{M_V}^2)(q^2-{M_V}^2)}
\times\nonumber\\
&&\left\{
{a-(p^2+q^2)(B_0F_0^2+b/{M_P}^2)\over r^2}
+{c+b/{M_P}^2(p^2+q^2)\over r^2-{M_P}^2} - B_0F_0^2 \right\}
\label{eq:vvpres3}
\end{eqnarray}
The parameters $a$, $b$ and $c$ may be related to properties
of the resonances which will feed back into the
low energy parameters which interest us upon expanding
the correlation function around $p^2=q^2=r^2=0$ and
comparing with the representation (\ref{eq:vvpdev}).
Here, we must exercise some care with respect to the
chiral loop part which is not explicitly displayed in
(\ref{eq:vvpdev}). Setting $p^2=q^2=0$ makes it
vanish\cite{dhl}\cite{bbc1}. This allows us to unambiguously
identify the parameter $a$ from the residue of the pion pole:
\begin{equation}
a={N_c\over 4\pi^2}B_0{M_V}^4
\label{eq:a}
\end{equation}
and to find a relation between $t_1$ and $c$:
\begin{equation}
32B_0t_1={-1\over2{M_V}^4}(B_0F_0^2 +{c\over{M_P}^2} )
\end{equation}
In the case of $t_2$, we must consider the terms proportional
to $p^2$. The chiral loop generates a contribution
proportional to $p^2/r^2$
(which cancels out the scale dependence
arising from $t_2$) and furthermore generates nonanalytic
terms which have no counterpart in the parametrization
(\ref{eq:vvpres2}). We will content ourselves
with a  simple estimate of $t_2$ valid in the
leading $N_c\ $ approximation: in this limit, the chiral
loop needs not be taken into account as it is subleading in $N_c\ $.
Ignoring the loop, one finds
\begin{equation}
4B_0t_2={1\over2{M_V}^4}( {a\over{M_V}^2} -B_0F_0^2- {b\over{M_P}^2} )
\end{equation}

There remains to find estimates for $b$ and $c$.
As usual, physical amplitudes are extracted from the Green's
functions by taking the residue of the appropriate resonance
poles and dividing out by the meson-current coupling constants.
Using that,
we can express $c$, to begin with, as a ratio of amplitudes:
\begin{equation}
c= a\tan\Theta {A(\pi'^0\to\gamma\gamma) \over
A(\pi^0\to\gamma\gamma)}
\end{equation}
where $\Theta$ is an angle which parametrizes the strength of the
coupling of the $\pi'$ to the pseudo-scalar current
\begin{equation}
<0\vert j^a_P(0)\vert\pi'^b>=\delta^{ab}B_0F_0\tan\Theta
\label{eq:tetadef}
\end{equation}
in the chiral limit
(the rationale for introducing an angle here is explained
in ref.\cite{ms}. )
As will be recalled in the next section (see (\ref{eq:l8sr}))
$\tan\Theta$ can
be related via a sum-rule to the LE parameter $L_8$.
The amplitudes $A(P\to\gamma\gamma)$
where $P$ is any pseudo-scalar meson are normalized
everywhere in the text such that the width:
\begin{equation}
\Gamma={1\over64\pi} M_P^3\, A^2
\label{eq:norm}
\end{equation}
Similarly, there is a simple relation between the parameter $b$
and the ratio of amplitudes $A(\pi\to\rho\gamma)/A(\pi\to\gamma
\gamma)$:
\begin{equation}
{A(\pi\to\rho\gamma)\over A(\pi\to\gamma\gamma)}
\left({-2e{F_V}\over{M_V}}\right)
= 1+x,\qquad
x=-{{M_V}^2\over a}(B_0F_0^2+ {b\over{M_P}^2} )
\label{eq:vmd}
\end{equation}
where ${F_V}$ is the coupling of the vector field to
the vector current. In eq.(\ref{eq:vmd}), $x$ measures the
deviation from the vector meson dominance
principle (VMD). Exact VMD means that $x=0$.
Using the experimental value \footnote{All the experimental
numbers are taken from the 1992 edition of the particle
data book \cite{pdb92} }
$\Gamma=68\pm7$ KeV for the decay width $\rho^+\to\pi^+\gamma$
(together with ${F_V}=150$ MeV and ${M_V}=770$ MeV) we obtain
$x=0.022\pm0.051$ which, as has long been well known, is
fairly close to exact VMD.  If we let ${M_P}$ go to infinity
in (\ref{eq:vmd}) we get a deviation
from VMD of the order of 20\%. For a better accuracy, we
have to use a finite $\pi'$ mass.
In this description, VMD is realized
from a cancellation between the parameter $b$ and the
asymptotic term $B_0F_0^2$ (see (\ref{eq:vmd}) ). Furthermore,
assuming the cancellation to be exact ensures that the
form factor associated with the matrix element
$<\gamma\vert j_\mu^a\vert\pi>$ satisfies an unsubtracted
dispersion relation.

We can now express the
parameters $t_1$ and $t_2$ in terms of experimentally
accessible resonance properties:
\begin{eqnarray}
&&t_1={-1\over64{M_V}^2}\left[ {F_0^2\over{M_V}^2} +{N_c\over4\pi^2}
\left({{M_V}\over{M_P}}\right)^2\tan\Theta\,
{A(\pi'\to\gamma\gamma)\over A(\pi\to\gamma\gamma)}
\right]\nonumber\\
&&t_2={N_c\over64\pi^2{M_V}^2}(1+x)
\label{eq:t1t2}
\end{eqnarray}
Since
the value of $x$ is compatible with zero,
the expression above for $t_2$
reproduces the one obtained previously\cite{pp} \cite{bbc2}.
Concerning $t_1$, the contribution proportional to
$A(\pi'\to2\gamma)$ is identical to the one which was identified
in ref.\cite{r+f} but the first contribution (which
comes from the asymptotic term in the three-point function) was
not included in that paper. We claim, on the contrary,
that this contribution is likely to dominate.
Experimentally,  there exists an upper bound for the decay
$\eta(1295)\to2\gamma$ rate,
$\Gamma<0.3$ KeV. Assuming ideal mixing in the $\pi(1300)$
multiplet together with the estimate $\tan\Theta\simeq2$
(which is obtained from the sum-rule quoted in sec.4
(\ref{eq:l8sr}) which relates it to the value of the LE
parameter $L_8$)
this bound implies that the first term in $t_1$
is larger than the second by a factor of at least three. This
means that keeping the first contribution in $t_1$, i.e.
ignoring the $\pi'$ contribution, should provide, at least,
the right sign and the right order of magnitude for this parameter.

Let us now consider the flavour singlet pseudo-scalar channel.
We can write a representation of the three-point function
analogous to (\ref{eq:vvpres3}):
\begin{eqnarray}
&&\Pi^0_{VVP}={1\over2(p^2-{M_V}^2)(q^2-{M_V}^2)}\times\nonumber\\
&&\left\{
{a'-(p^2+q^2)(B_0F_0^2+b'/{M_{P'}}^2)\over r^2-M^2_{\eta'}}
+{c'+b'/{M_{P'}}^2(p^2+q^2)\over r^2-{M_{P'}}^2}
-B_0F_0^2 \right\}
\label{eq:pi0vvp}
\end{eqnarray}
Here, ${M_{P'}}$ is the mass of the singlet state
in the $\pi(1300)$ nonet in the chiral limit.
At leading $N_c\ $ order we expect
nonet symmetry to hold and thus ${M_{P'}}={M_P}$. Furthermore, in
this limit, the residues of the poles in (\ref{eq:pi0vvp}) should
be equal to the ones in (\ref{eq:vvpres3}), i.e $a=a'$, $b=b'$
$c=c'$.
One can express $t'_1$ as a difference between $\Pi_{VVP}$
and $\Pi^0_{VVP}$ in the limit
$p^2=q^2=r^2=0$ (again the loop contribution vanishes):
\begin{equation}
96B_0t'_1=\lim_{r^2\to0}\Big[
\Pi^0_{VVP}(0,0,r^2)-\Pi_{VVP}(0,0,r^2)+{B_0N_c\over8\pi^2 r^2}\Big]
\end{equation}
which gives
\begin{equation}
96B_0t'_1={-1\over2{M_V}^4}\Big({a'\over{M_{\eta'}}^2}+{c'\over{M_{P'}}^2}
-{c\over{M_P}^2}\Big)
\label{eq:tprime}
\end{equation}
Here, one expects a strong cancellation between the $c$ and the
$c'$ term. In fact, the first term in the parenthesis
is of order $O(N^2_c)$
while the sum of the last two terms is of order
$N_c^0$. It is easy to check that adding further resonances
would also generate corrections which are of order
$N_c^0$ and which are suppressed by inverse powers of the resonance
masses. Therefore, it appears that retaining only the first term
in (\ref{eq:tprime}) should constitute a rather solid approximation.
Next, using (\ref{eq:pi0vvp})
we can express the parameter $a'$ in terms of the amplitude
$A({\eta'}\to2\gamma)$. Introducing the coupling of the
${\eta'}$ to the pseudo-scalar singlet current
\begin{equation}
<0\vert j^0_P(0)\vert{\eta'}>=B_0G_{\eta'}
\label{eq:getap}
\end{equation}
we obtain the following representation for the
parameter $t'_1$.
\begin{equation}
e^2t'_1={-G_{\eta'}\over256{M_{\eta'}}^2}\sqrt6 \,
A({\eta'}\to\gamma\gamma)
\label{eq:t'1}
\end{equation}
(the amplitude is again normalized as in (\ref{eq:norm}) )
At this point we have all the necessary ingredients to
discuss the chiral corrections to the decays
$\eta\to\gamma\gamma$ and $\pi^0\to\gamma\gamma$.

\vspace{2mm}
\centerline{\bf 3. APPLICATION TO $\pi^0$ AND $\eta$ DECAYS}
\vspace{2mm}

Let us begin with the $\eta$. Collecting the chiral lagrangian
pieces up to order $O(p^6)$ one obtains:
\begin{equation}
{\cal L}_{\phi_8\to2\gamma}={e^2F_{\mu\nu} \tilde F^{\mu\nu}\phi_8\over
16\pi^2 F_0\sqrt{3} } \left\{
1 +{5-2r_2\over3}\, T_1
  +(1-r_2)\,T'_1\right\}
\label{eq:lageta}
\end{equation}
where we have introduced for convenience the dimensionless
quantities proportional to $t_1$ and $t'_1$:
\begin{equation}
T_1={256\pi^2\over3}{M_\pi}^2 t_1\quad\quad
T'_1={1024\pi^2\over3}{M_\pi}^2 t'_1
\end{equation}
In (\ref{eq:lageta}) $r_2$ is the quark mass ratio $m_s/{\hat m}$ which
must be expressed at $O(p^2)$ precision, i.e.
\begin{equation}
r_2=2M_K^2/M_\pi^2-1\simeq 25.9
\label{eq:r2}
\end{equation}
and, as usual $\tilde F^{\mu\nu}=1/2\epsilon^{\mu\nu\alpha\beta}
F_{\alpha\beta}$.
In order to compute the $\eta$ decay amplitude from (\ref{eq:lageta})
we must use that $\phi_8=\phi_\eta$ at chiral order $O(p^2)$ while
$\phi_8=\phi_\eta F_0/F_\eta$ at order $O(p^4)$, ignoring the
$\pi^0-\eta$ mixing which effect can be shown to be negligible
for the $\eta$ decay.

Since the loop contributions vanish for on-shell photons \cite{dhl}
\cite{bbc1} the decay rate is computed by simply
using (\ref{eq:lageta})
at tree level. Using relation (\ref{eq:t'1}) for $t'_1$ one can
write the amplitude for $\eta$ decay in the following way
\begin{equation}
A(\eta\to\gamma\gamma)={\alpha\over\sqrt3\pi{F_\pi}}\left\{
{{F_\pi}\over{F_\eta}} + {5-2r_2\over3}\,  T_1\right\}
+{\sqrt2\over3}(r_2-1){{M_\pi}^2\over{M_{\eta'}}^2}{{G_{\eta'}}
\over{F_\pi} }A({\eta'}\to\gamma\gamma)
\label{eq:aeta}
\end{equation}
In this formula, we have replaced $F_0$ everywhere by ${F_\pi}$:
this does not modify the first term, and in the others,
this replacement amounts to a modification of order $O(p^8)$.
One observes that the $\eta'$ amplitude has crept in
via the sum-rule formula for the LE parameter $t'_1$.

Using
this form of the amplitude we can make contact with the
traditional parametrization of the $\eta$ and $\eta'$ decay
amplitudes in terms of an $\eta-\eta'$ mixing angle. This
will allow us to establish definite relations between the
parameters appearing in this representation and those
of the chiral expansion. The representation which is
usually employed contains three parameters $\theta_0$,
$\lambda_0$ and $\lambda_8$ (see e.g. \cite{gk})
\begin{eqnarray}
&&A(\eta\to\gamma\gamma)={\alpha\over\sqrt3\pi{F_\pi}}
\Big( {\cos\theta_0\over\lambda_8}
-{\sqrt8\sin\theta_0\over\lambda_0}\Big) \nonumber\\
&&A({\eta'}\to\gamma\gamma)={\alpha\over\sqrt3\pi{F_\pi}}
\Big( {\sin\theta_0\over\lambda_8}
+{\sqrt8\cos\theta_0\over\lambda_0}\Big)
\end{eqnarray}
Let us eliminate $\lambda_0$ from the second equation and
replace in the first one. We obtain a form exactly similar
to (\ref{eq:aeta}), which allows us to express
$\lambda_8$ and $\theta_0$ in terms of LE parameters:
\begin{equation}
{1\over\cos\theta_0\lambda_8}={{F_\pi}\over{F_\eta}}+
{5-2r_2\over3}\, T_1
\end{equation}
The usual assumption made in the literature to identify
$\lambda_8$ with the ratio ${F_\eta}/{F_\pi}$ is seen to be
justified only to the extent that the correction from $t_1$
is negligible. Here, we find that this correction is of the order
of 10\%.  Concerning the mixing angle, we obtain
\begin{equation}
\tan\theta_0={-\sqrt{2}\over3}(r_2-1){{M_\pi}^2\over{M_{\eta'}}^2}
\,{{G_{\eta'}}\over{F_\pi}}
\label{eq:chimix}
\end{equation}
A relation between the mixing angle and the parameter $L_7$
was first derived in \cite{gl85} based on an analysis of
the pseudoscalar mass formulas.
We obtain a similar, but slightly different, formula upon
using the sum-rule relation between $G_{\eta'}$ and $L_7$
(see (\ref{eq:l7sr}) below).
The two formulas do coincide at first order in the
quark masses, as they should.

Let us now make some numerical estimates. At order $O(p^4)$,
first, one has the well-known result
\begin{equation}
\Gamma^{(4)}_{\eta\to2\gamma}={\alpha^2{M_\eta}^3\over192\pi^3{F_\pi}^2}
\simeq172\ {\rm eV}
\end{equation}
(In numerical applications we use the value
${F_\pi}=92.4\pm0.2$ MeV\cite{fpi}).
At order $O(p^6)$ now, we can use ${F_\eta}=(1.3\pm0.05){F_\pi}$
provided by \cite{gl85} and  $t_1$ is given by (\ref{eq:t1t2}) and
the following discussion. In (\ref{eq:aeta}) we still need to evaluate
the coupling constant $G_{\eta'}$. An estimate may be obtained
from the sum-rule for the LE parameter $L_7$\cite{gl90}
\begin{equation}
L_7={-{G_{\eta'}}^2\over48{M_{\eta'}}^2}
\label{eq:l7sr}
\end{equation}
This result is derived in the chiral limit under the only assumption
that the ${\eta'}$ pole is the dominant contribution in the
dispersion relation. Using the numerical value of $L_7$,
$L_7=-(0.4\pm0.15)\,10^{-3}$ \cite{gl85} one deduces that
${G_{\eta'}}=133\pm25$ MeV up to a sign. In the large $N_c\ $ limit
we expect to have ${G_{\eta'}}={F_\pi}$ which incites us to adopt the
positive sign.
Using this value in the
expression (\ref{eq:aeta}) for the amplitude, together with the
experimental value of the width $\Gamma({\eta'}\to\gamma\gamma)=
4.29\pm0.19$ KeV, we obtain the $\eta$ width
in the following form
\begin{equation}
\Gamma(\eta\to2\gamma)=172\,\Big[(0.77\pm0.03)+ 0.09
+(0.72\pm0.15)\Big]^2
=430\pm98\ {\rm eV}
\end{equation}
where we have displayed the contributions in the same order
as they appear in (\ref{eq:aeta}).
One observes that the third contribution,
which comes from the parameter $t'_1$ largely dominates
over the one from $t_1$, in agreement with the expectation
from large $N_c\ $ arguments. The contribution from $t_1$ is not
completely negligible, it increases the result by about
50 eV.
If we had chosen
the opposite sign for ${G_{\eta'}}$
(implying a huge deviation from the leading $N_c\ $
estimate) then the width would have been of the order of 3 eV.
The central value is slightly below
the experimental result (the average of the two-photon
data at present give $\Gamma=510\pm26$ eV ) but
the relatively large error bar on $L_7$
generates an uncertainty of nearly 100 eV on our prediction (as a
matter of fact, within this uncertainty, the result is compatible
with both the old Primakov result ($\Gamma=324\pm46$ eV) and
the photon-photon results).  The uncertainty on $L_7$ is
tightly correlated with the uncertainty on the value of the
quark mass ratio $r$, which is $r=25.7\pm2.3$ in the standard
$\chi$pt at $O(p^4)$\cite{gl85}.
Roughly speaking, the upper bound on $\Gamma$ corresponds to the
lower bound on $r$ and vice versa.

Using our formula (\ref{eq:chimix}) with ${G_{\eta'}}$ derived from
$L_7$, we obtain for the mixing angle
\begin{equation}
\theta_0=-(18.4\pm3.6)^\circ
\end{equation}
which is smaller than the one quoted in \cite{gl90},
$\theta_0=-(22\pm4)^\circ$.
We should note here that the value of $L_7$ that we used
does not take into account the recent work which have provided
estimates for the corrections to Dashen's theorem\cite{urech}.
Part of the large error bar in $L_7$ is generated from the
uncertainty associated with these corrections.

Let us now turn to the $\pi^0$ decay. In contrast to the $\eta$
case it is crucial here to take the quark mass difference $m_d-m_u$
into account. The
lagrangian including the $O(p^6)$ corrections reads
\begin{equation}
{\cal L}_{\phi_3\to\gamma\gamma}=
{e^2F_{\mu\nu} \tilde F^{\mu\nu}\phi_3\over
16\pi^2 F_0} \left\{ 1 +
(1-{5\over3}{m_d-m_u\over m_d+m_u} )\, T_1
+{m_u-m_d\over m_u+m_d}\, T'_1
\right\}
\end{equation}
Clearly, one expects the chiral corrections here to be much smaller
than in the $\eta$ case because they do not involve the strange
quark mass. The wave-function renormalization, in this case
can be written as follows at $O(p^4)$\cite{gl85} neglecting terms
which are quadratic in $\epsilon$
\begin{equation}
\left(
\begin{array}{c}
\phi_3\\
\phi_8
\end{array}
\right)
=
\left(
\begin{array}{cc}
1 &-\epsilon_1\\
\epsilon_2&  1
\end{array}
\right)
\left(
\begin{array}{cc}
F_0/F_{\pi^0}&0\\
0                &  F_0/F_\eta
\end{array}
\right)
\left(
\begin{array}{c}
\phi_\pi\\
\phi_\eta
\end{array}
\right)
\end{equation}
At $O(p^2)$ one has to set $\epsilon_1=\epsilon_2=\epsilon$ with
\begin{equation}
\epsilon={\sqrt3\over4}{m_d-m_u\over m_s-{\hat m}}
\label{eq:eps}
\end{equation}
and, numerically, $\epsilon\simeq1.00\,10^{-2}$ and
$\epsilon_2\simeq1.11\,10^{-2}$.
Expressing next $t'_1$ in terms of the mixing angle $\theta_0$
and the ${\eta'}$ decay amplitude as before, we
can write the amplitude for the pion decay in the following
form:
\begin{equation}
A(\pi\to\gamma\gamma)={\alpha\over\pi F_{\pi^0}}\left\{
1 + {\epsilon_2\over\sqrt3}+
\Big(1+{5-4r_2\over\sqrt3}\epsilon\Big)T_1
\right\}
-3\epsilon\,\tan\theta_0\,A({\eta'}\to\gamma\gamma)
\end{equation}
One notices that $F_{\pi^0}$ appears
here and not ${F_\pi}\equiv F_{\pi^+}$
which is experimentally well determined. Theoretically, there are
two contributions to the difference between $F_{\pi^0}$ and
$F_{\pi^+}$. One arises from the quark mass difference $m_d-m_u$:
a simple estimate (neglecting chiral logarithms) gives
\begin{equation}
F_{\pi^0}=F_{\pi^+}  \Big(
1-{2\over3}({{F_K}^2\over{F_\pi}^2}-1)\epsilon^2 +{1\over2}\epsilon_2^2
-\epsilon_1\epsilon_2\Big)
\end{equation}
which shows that this contribution is negligibly small, of the
order of $10^{-4}$. The other contribution is purely
electromagnetic and, being of order $O(\alpha_{QED})\simeq
10^{-2}$ it should be the dominant one. Unfortunately,
no estimate of this effect seems to be available at present.

Using the value of $\theta_0$
as discussed above, one can express the rate including the $O(p^6)$
corrections as
\begin{equation}
{F^2_{\pi^0}\over F^2_{\pi^+}}\Gamma=
7.73\,\Big[1+(6.41-2.49+12.5)10^{-3}\Big]^2=7.98\pm0.08\,{\rm eV}
\label{eq:pilarg}
\end{equation}
In the parenthesis we have displayed the contributions
from the terms proportional to $\epsilon_2$ , $T_1$ and $\tan\theta_0$
respectively. The value of the error includes the uncertainty on
${F_\pi}$ as well as that from $L_7$.
Comparison with the experimental result, $\Gamma=7.74\pm0.60$ eV,
suggests that the value of $F_{\pi^0}$ should be somewhat larger
than ${F_\pi}$. Despite the rather large error bar on the $\pi^0$ width,
the value quoted in ref.\cite{pdb92}, $F_{\pi^0}=84\pm3$ MeV, is
incompatible with our chiral expansion of the amplitude.

Let us finally discuss the description of the amplitudes
$\eta(\pi^0)\to\gamma\gamma$ in the generalized $\chi$pt approach. The
$O(p^4)$ lagrangian in the Wess-Zumino sector is obviously the same
as in the conventional $\chi$pt, being independent of the quark masses.
The leading corrections are given by exactly the same two terms,
proportional to $t_1$ as $t'_1$ as before,
which are counted as $O(p^5)$ rather than $O(p^6)$.
The main difference then, is that the value of the quark mass
ratio $r$ is no longer determined from the $O(p^2)$ $\chi$pt but is
left as a free parameter. As a function of this parameter, we can
express the $\eta$ amplitude as follows:
\begin{equation}
A(\eta\to\gamma\gamma)={\alpha\over\sqrt3\pi{F_\pi}}\left\{
{{F_\pi}\over{F_\eta}}+{5-2r\over3}\,T_1(r)\right\}
-\tan\theta_0(r) A({\eta'}\to\gamma\gamma)
\end{equation}
Concerning the $\pi^0$, we note first that
the mixing parameters depend on $r$,
for instance
\begin{equation}
\epsilon(r)=\epsilon\,{1\over r_2-1+\Delta_{GMO}/2}\Big(
r-1+{2r(r_2-r)\over r+1}-\Delta_{GMO} \Big)
\end{equation}
where $\Delta_{GMO}=3{M_\eta}^2/{M_\pi}^2-2r_2-1$
and $\epsilon$ has the same meaning as before (\ref{eq:eps}). The amplitude
for $\pi^0$ decay can be written as
\begin{eqnarray}
A(\pi\to\gamma\gamma)&&={\alpha\over\pi F_{\pi^0}}\left\{
1+{\epsilon_2(r)\over\sqrt3}+
\Big[1+{10(1-r)\over3\sqrt3}\epsilon
+{5-2r\over3\sqrt3}\epsilon(r)\Big]\,T_1(r)
\right\}                                          \nonumber\\
&&-[2\epsilon+\epsilon(r)]\,\tan\theta_0(r)\,A({\eta'}\to\gamma\gamma)
\end{eqnarray}
The expression for $\epsilon_2(r)$ is rather lengthy and contains
several $O(p^3)$ low-energy parameters which are not precisely known,
however, an approximate expression can be derived
\begin{equation}
\epsilon_2(r)=\epsilon(r)+\epsilon\,{2\over3}(2r+1){r-r_2\over r^2-1}
\,\Big({{F_K}^2\over{F_\pi}^2}-1\Big)
\end{equation}

The determination of the mixing angle angle $\theta_0(r)$ goes
in the same way as before. Instead of eq.(\ref{eq:chimix}), we obtain
here
\begin{equation}
\tan\theta_0(r)={\sqrt2\over3}(r-1){2{\hat m} B_0\over{M_{\eta'}}^2}
{{G_{\eta'}}\over{F_\pi}}=
{1\over\sqrt3}{{M_\pi}\over{M_{\eta'}}}\left(
-\Delta_{GMO} +{2(r_2-r)(r-1)\over(r+1)} \right)^{1\over2}
\label{eq:chimixr}
\end{equation}
The second equality is obtained upon using the sum-rule
(\ref{eq:l7sr}). Instead of $L_7$, one must use the corresponding LE
parameter $Z_0^P$ in the generalized $\chi$pt,
which appears at $O(p^2)$ and
is given at this order from the deviation to the Gell-Mann-Okubo
mass formula (see \cite{daphne}). In order to determine
$T_1(r)$ one starts, as before,  from the sum-rule
(\ref{eq:t1t2}) for $t_1$. The coupling $\tan\Theta$ is estimated by
using, again, the sum-rule (\ref{eq:l8sr}) replacing $L_8$ by the
corresponding parameter, $A_0$, in the G$\chi$pt. One ends up with the
following expression:
\begin{equation}
T_1(r)={-4\pi^2\over3}{{M_\pi}^2{F_\pi}^2\over{M_V}^4}(1-\lambda)
-{{M_\pi}\over{M_P}}\left({\lambda{M_S}^2-(1-\lambda)^2{M_\pi}^2\over
{M_P}^2-{M_S}^2}\right)^{1\over2} {A(\pi'\to\gamma\gamma)\over
A(\pi\to\gamma\gamma)},\quad \lambda={2(r_2-r)\over r^2-1}
\end{equation}
Let us now consider some numerical results in this formalism
for a small value of $r$, $r=10$ for example. Using the
experimental bound on $\eta(1295)\to\gamma\gamma$ we find
\begin{equation}
T_1(r=10)=-6\,10^{-3} (0.68 +X ),\qquad \vert X\vert<1.38
\end{equation}
where $X$ is the term proportional to $A(\pi'\to\gamma\gamma)$
in the formula for $T_1(r)$. This term generates an uncertainty
of the order of 30 eV in the value of the $\eta$ width and
of the order of 0.1 eV in the $\pi^0$ width. The
remaining contributions to the $\eta$ width are shown below:
\begin{equation}
\Gamma(\eta\to\gamma\gamma)\vert_{r=10}=172 \, \Big[
0.77 +0.02 +0.95\Big]^2=521\,{\rm eV}
\end{equation}
where we have used that ${F_\eta}/{F_\pi}$ is practically the same as
in the standard $\chi$pt and that the value of the mixing angle
for $r=10$, as obtained from (\ref{eq:chimixr}),
is $\theta_0\simeq23.8^\circ$. This result is at the upper limit
of the allowed range in the standard $\chi$pt. The various factors
in (\ref{eq:chimixr}) are, however,
widely different from the corresponding ones
in (\ref{eq:chimix}): $r<<r_2$, $2{\hat m} B_0 <<{M_\pi}^2$,
but this is compensated
by the third factor, $G_{\eta'}\simeq 8{F_\pi}$ which is much larger
than in the standard $\chi$pt (and deviates more from the large $N_c\ $
result).
The $\pi^0$ width,
finally, can be expressed as
\begin{equation}
\Gamma(\pi^0\to\gamma\gamma)\vert_{r=10}=8.07\,
{F^2_{\pi^+}\over F^2_{\pi^0}}\,{\rm eV}
\end{equation}
Both the results for the $\pi^0$ and the $\eta$ fall in the range of
values which are also allowed in the standard $\chi$pt.

\vspace{2mm}
\centerline{\bf 4. MIXED LARGE $N_c\ $ AND CHIRAL EXPANSION}
\vspace{2mm}

The proper way to introduce the pseudo-scalar singlet
field $\phi_0(x)$ into the
chiral lagrangian was first discussed in
\cite{divecchia}. The most general lagrangian at order $O(p^2)$
is discussed in \cite{gl85}.
We will follow their method here, which consists in introducing
a source $\theta(x)$ associated with the winding number density
$\omega=(16\pi^2)^{-1}{\rm Tr}G_{\mu\nu}\tilde G^{\mu\nu}$.
Under an axial U(1) transformation, $\exp(i\beta(x))$, one lets
the source $\theta(x)$ transform to $\theta(x)-2{\rm tr}\beta(x)$. Then,
(apart from the canonical WZ part) the effective action is
invariant. In the singlet sector, three
kinds of invariant building blocks are available
$D_\mu\phi_0=\partial_\mu
\phi_0-2{\rm tr} a_\mu$, $D_\mu\theta=
\partial_\mu\theta+2{\rm tr} a_\mu$ and
the combination $\sqrt6\phi_0/F_0+\theta$.
For the purpose of
discussing the spectrum and electromagnetic properties it is
sufficient to consider quadratic polynomials
in $\phi_0(x)$. The canonical WZ action is minimally extended
using nonet symmetry.

At order two the mixing angle is predicted to be
$\theta_P\simeq-10^\circ$\cite{dhl}\cite{gl85} which is too
small by a factor of two compared to phenomenological determinations
\cite{gk}.
It has long been recognized that
going to $O(p^4)$ seems to improve the situation\cite{dhl}. In
ref. \cite{gl85} it is proposed to use the $O(p^4)$ lagrangian with
the ${\eta'}$ integrated out in conjunction with the $O(p^2)$ lagrangian
which includes the ${\eta'}$. Although this procedure gives the
correct magnitude for $\theta_P$ one may wish to include the ${\eta'}$
at $O(p^4)$ as well.
Furthermore, the lagrangian proposed in \cite{gl85}
contains terms which are leading in $N_c\ $ as well as terms which are
subleading. Outside of the large $N_c\ $ expansion,
including the ${\eta'}$ explicitly into the chiral lagrangian
does not make much sense since the ${\eta'}$ mass is rather
large, larger than the mass of the $\rho$ or $\omega$ mesons,
for instance.
On the contrary, this looks reasonable if one thinks in terms of the
large $N_c\ $ expansion, since $M^2_{\eta'}$ is $O$(1/$N_c\ $).
This suggests to generalize the chiral
counting in order to include the large $N_c\ $ expansion:
a natural choice is to count
a factor which is $O(1/N_c)$ on the same footing as a quark mass
factor\cite{stern}.
We will call this counting rule the mixed chiral expansion. It
should be clear that this expansion is different from the
usual chiral expansion. For instance, at leading order,
the eta mass does not satisfy
the Gell-Mann-Okubo relation and the ratio ${F_\pi}/{F_\eta}$ differs
from one. However, it is also
a systematic expansion which could, in principle, be carried out
at arbitrary high order.

We first generalize the chiral matrix $U$ to include a nonet
of fields:
\begin{equation}
\bar U = U \exp i\Big( {\lambda_0 \phi_0(x)\over F_0}\Big)
\end{equation}
At order two of the mixed expansion one has the two terms
which are $O(p^2)\times O(N_c)$ as well as a single term which is
$O(p^0)\times O( (N_c)^0)$:
\begin{equation}
\bar{\cal L}_{(2)}={F_0^2\over4}\left\{
<D_\mu\bar U^\dagger D^\mu\bar U +
\chi^\dagger \bar U +\bar U^\dagger  \chi>
-{1\over3}M_0^2 ( \sqrt6{\phi_0\over F_0}+\theta)^2\right\}
\label{eq:ll2}
\end{equation}
It is not difficult to extend this to the next order.
We note first that
chiral loops need not be considered since they are
of chiral order four and they are suppressed by one power of
$N_c\ $,
so we effectively count this kind of
contribution as order six. Furthermore, the two terms of the standard
$O(p^4)$ lagrangian:
\begin{equation}
{\cal L}=L_6<\chi^\dagger\bar U +\bar U^\dagger \chi>^2
+    \hat L_7<\chi^\dagger\bar U -\bar U^\dagger \chi>^2
\end{equation}
are {\sl both} suppressed by the Zweig rule here so they are also
effectively of order six (the $L$'s pick up contributions
from the resonances which are {\sl not} included in the chiral
lagrangian, this is why $\hat L_7$ is different from $L_7$ ). The
only terms that we need consider are
\begin{equation}
\bar{\cal L}_{(4)}=L_5<
D_\mu\bar U^\dagger D^\mu\bar U
(\chi^\dagger \bar U +\bar U^\dagger  \chi)>
+L_8< \chi^\dagger \bar U \chi^\dagger \bar U
+     \bar U^\dagger  \chi\bar U^\dagger  \chi>
\end{equation}
In addition to these terms, which are $O(p^4)\times O(N_c)$, we have
to consider the terms which are $O(p^2)\times O((N_c)^0)$.
We can form two terms of this kind:
\begin{equation}
{\cal L}'_{(4)}=k_1 D_\mu\phi_0 D^\mu\phi_0
+i k_2{F^2_0\over6}(\sqrt6{\phi_0\over F_0}+\theta)
<\chi^\dagger\bar U -\bar U^\dagger  \chi>
\label{eq:ll4}
\end{equation}
Furthermore, in the Wess-Zumino sector,
we can form a term which is $O(p^4)\times O((N_c)^0)$ and is
thus on a similar footing with the true $O(p^6)$ corrections:
\begin{equation}
{\cal L}'^{WZ}_{(6)}=k_3 \epsilon^{\mu\nu\alpha\beta} <f^{(+)}_{\mu\nu}
f^{(+)}_{\alpha\beta}>(\sqrt6{\phi_0\over F_0}+\theta)
\end{equation}

Before proceeding, we note that the LE
parameters $L_5$, $L_8$ and the quark mass ratio $r=m_s/{\hat m}$
have to be re-determined using the logic of the mixed expansion
scheme adopted here. In particular, $L_5$ and $L_8$ are
scale independent at order four now. Outside of the
$\eta-\eta'$ sector one may simply use
the formulae of \cite{gl85} and drop the loop contributions. $L_5$ ,
to begin with, is
related to the ratio $F_K/F_\pi\simeq1.22$:
\begin{equation}
L_5={F^2_K-F^2_\pi\over8(M^2_K-M^2_\pi)}\simeq2.31\,10^{-3}
\end{equation}
This value is very close to the result of \cite{gl85}
$L_5(\mu=M_\eta)=2.3 10^{-3}$ which is not very surprising,
since the kaon
and the eta chiral logarithms are very small for this value
of the scale and the pion chiral logarithm, being proportional
to the pion mass squared, is small anyway.

Next, expressing the $K$ and $\pi$ masses
\begin{eqnarray}
&&(1+{8M^2_\pi\over F^2_\pi }L_5) M^2_\pi=2{\hat m} B_0 +
           {16M^4_\pi\over F^2_\pi }L_8\nonumber \\
&&(1+{8M^2_K\over F^2_\pi }L_5) M^2_K=(r+1){\hat m} B_0 +
           {16M^4_K\over F^2_\pi }L_8
\end{eqnarray}
we can solve for ${\hat m} B_0$ and $L_8$. For the latter, we find
\begin{equation}
L_8={L_5\over2}+{F^2_\pi\over8M^2_\pi}\,{r_2-r\over r_2^2-1}
\label{eq:l8val}
\end{equation}
In order to evaluate $L_8$, we need to know the value of $r$.
If we use the number given in \cite{gl85}, $r=25.7$ (with
a 10\% uncertainty), we find $L_8=1.17\,10^{-3}$ (again
fairly close to $L_8({M_\eta})=1.1\,10^{-3}$). However, recent
work on the $O(e^2 m_q)$ corrections to Dashen's theorem
may imply that this value of $r$ should be revised downwards.
Indeed, \cite{gl85} have shown that $r$ is related to the strong part
of the $K^0-K^+$ mass difference by the following relation
\begin{equation}
r+1 =(r_2+1) {m_d-m_u\over m_s-{\hat m}} \, {M^2_K - M^2_\pi\over
(M^2_{K^0} -M^2_{K^+})_{QCD} }
\label{eq:petitr}
\end{equation}
In this relation, \cite{gl85} argue that
$R=(m_s-{\hat m})/(m_d-m_u)$ can be estimated from the baryon sector
or from $\omega-\rho$ mixing
in a consistent way, giving\cite{gl82}
$R=43.5\pm3.2$.
The strong
interaction part of the $K_0 - K_+$ mass difference  can be written
as
\begin{equation}
( M^2_{K^0} - M^2_{K^+})_{QCD} = M^2_{K^0} -M^2_{K^+}
+M^2_{\pi^+} -M^2_{\pi^0} +\delta (M^2_{\pi^+} -M^2_{\pi^0} )
\end{equation}
Dashen's theorem\cite{dashen} states that
$\delta$ is a quantity which is $O(e^2m_q)$.
Recent estimates \cite{urech} yield $\delta$ to
be positive and in the range
$0.4\le\delta\le0.8$. In the apparent absence of any good reason
to doubt the estimate of $R$, using this in
relation (\ref{eq:petitr}) implies that the central value
of $r$ should be moved down to $r\simeq23$. This has the consequence
of increasing the value of $L_8$ up to $L_8=1.43\,10^{-3}$. This
innocent looking modification has a strong influence
on the eta decay, as we will see.

It is interesting to compare these results with sum-rule estimates.
For $L_5$ \cite{gl90} gives
\begin{equation}
L_5={F_0^2\over4M^2_S}\simeq2.3 10^{-3}
\end{equation}
(using $M_S=M_{a_0}=980$ MeV )
which is in fairly good agreement with the above evaluation. A less
publicized chiral sum-rule can also
be derived\cite{ms} giving $L_8$ in the following form:
\begin{equation}
L_8={F^2_0\over16M^2_S}\left(1 +\tan^2\Theta(1-{M^2_S\over M^2_P})
\right)
\label{eq:l8sr}
\end{equation}
where $\Theta$ is defined in (\ref{eq:tetadef}).
Using the value of $L_8$ found above
one obtains the estimate of $\Theta$ used in sec.2.

Let us now consider the masses of the $\eta$ and the ${\eta'}$. At
order four we have not only a mass matrix {\bf M}
but also a
kinetic matrix {\bf K}.  Both matrices are real and symmetric and
the matrix elements of the mass matrix are
\begin{eqnarray}
&& M_{11}= {1\over3}{M_\pi}^2\left[
2r_2+1+2r_2(r_2+2)\,z+4(r_2-1)^2\,y
\right]\nonumber\\
&&M_{12}={-\sqrt2\over3}(r_2-1){M_\pi}^2\Big[
1+(r_2+3)\,z+2(r_2-1)\,y +2k_2-k_1
\Big]   \nonumber\\
&&M_{22}=M_0^2(1-2k_1) +{1\over3}{M_\pi}^2
\left[r_2+2+((r_2+1)^2+2)\,z +2(r_2-1)^2\,y
+2(r_2+2)(2k_2-k_1)  \right]\nonumber\\
\end{eqnarray}
where \begin{equation}
y=4M^2_\pi L_8/F^2_\pi,\  \qquad z=4M^2_\pi L_5/F^2_\pi
\end{equation}
The mixing with the $\pi^0$
has a negligibly small effect at this level and will be ignored.
The matrix elements of the kinetic matrix now are
\begin{eqnarray}
K_{11}=1+{2\over3}(2r_2+1)z,\quad K_{12}=-{2\sqrt{2}\over3}(r_2-1)z,
\quad K_{22}=1+{2\over3}(r_2+2)z
\end{eqnarray}
The mass squared of the $\eta$ and
${\eta'}$ mesons are given by the eigenvalues of the generalized
eigenvalue problem:
\begin{equation}
\left(
\begin{array}{cc}
M_{11}&M_{12}\\
M_{12}&M_{22}
\end{array}
\right)
\left(
\begin{array}{cc}
V_{11}&V_{12}\\
V_{21}&V_{22}
\end{array}
\right)
=
\left(
\begin{array}{cc}
K_{11}&K_{12}\\
K_{12}&K_{22}
\end{array}
\right)
\left(
\begin{array}{cc}
V_{11}&V_{12}\\
V_{21}&V_{22}
\end{array}
\right)
\left(
\begin{array}{cc}
{M_\eta}^2& 0 \\
 0     & {M_{\eta'}}^2
\end{array}
\right)
\end{equation}
As a result, the mixing matrix {\bf V}
is not unitary (instead of satisfying $^t${\bf VV}=1 it satisfies
$^t${\bf VKV}=1). This implies that one has to introduce
two mixing angles $\theta_0$
and $\theta_8$ and not just one as is usually done
(this fact was pointed out in ref.\cite{schechter}):
\begin{eqnarray}
&&\phi_8={1\over\lambda_8}
(\phi_\eta\cos\theta_8 +\phi_{{\eta'}}\sin\theta_8)  \nonumber\\
&&\phi_0={1\over\lambda_0}
(-\phi_\eta\sin\theta_0 +\phi_{{\eta'}}\cos\theta_0 )
\label{eq:mixing}
\end{eqnarray}
The three parameters $k_1,\ k_2$ and $M_0$ which appear
in the mass matrix can effectively be absorbed into two
unknowns $M_0^2(1-2k_1)$ and $2k_2-k_1$, which we expect
to determine from fitting the $\eta$ and ${\eta'}$ masses.
It turns out that one has two different solutions, which
we label as $(+)$ and $(-)$.
The requirement that the large $N_c\ $ expansion be meaningful allows one
to eliminate one of the solutions. Indeed, numerically, the solution
(+) corresponds to the following values for the parameters $M_0$ and
$k_2$ (setting $k_1=0$):
\begin{equation}
M_0=898 \ {\rm MeV}\qquad k_2=0.12
\end{equation}
while the solution $(-)$ corresponds to
\begin{equation}
M_0=1226 \ {\rm MeV}\qquad k_2=-1.43
\end{equation}
Clearly, the solution (+) does satisfy the criterion that the
contribution of the $k_2$ term in the mass matrix is roughly
of the same magnitude as, say, the contribution from the $L_8$
term. This does not hold for the other solution. The possibility
of reproducing exactly the $\eta$ and ${\eta'}$ masses while
having only small corrections to the $O(p^2)$
lagrangian (\ref{eq:ll2}) was noted by \cite{peris}.
One can express the mixing angles in closed form
\begin{equation}
\tan\theta_0=-\left({M_{11}-K_{11}{M_\eta}^2\over
K_{11}{M_{\eta'}}^2-M_{11}}\right)^{1\over2}, \qquad
\theta_8^\pm=\theta_0\pm\phi,\qquad
\tan\phi={-K_{12}\over (K_{11}K_{22}-K_{12}^2)^{1\over2}}
\label{eq:mix3}
\end{equation}
as well as the extra factors
\begin{equation}
\lambda_8=(K_{11}-{K^2_{12}\over K_{22}})^{1\over2} \qquad
\lambda_0=(K_{22}-{K^2_{12}\over K_{11}})^{1\over2}
\label{eq:mix4}
\end{equation}
In practice, however, these expressions should be expanded linearly
in terms of the $O(p^4)$ parameters in order to remain consistent with
the $O(p^4)$ precision.
Before exploiting this, let us examine the results
that one would obtain at order two of the mixed expansion,
i.e. starting from (\ref{eq:ll2}).
The unique parameter in that case,
$M_0$, may be adjusted so as to reproduce the $\eta$ mass.
The
mixing angle and the $\eta'$ mass then satisfy
\begin{equation}
\tan\theta_P=-{(2r_2+1){M_\pi}^2-3{M_\eta}^2\over\sqrt{2}(r_2-1){M_\pi}^2},
\quad
{\hbox{$M^2_{\eta'}$\kern-1.25em \raise 1.7ex\hbox{$\circ$}\kern.7em}}
={M_\pi}^2\left( {2r_2+1\over3}+
{\sqrt2(r_2-1)\over3\tan\theta_P}\right)
\end{equation}
Numerically, this gives
${\hbox{$M_{\eta'}$\kern-1.25em \raise 1.7ex\hbox{$\circ$}\kern.7em}}
=1583$ MeV, which is indeed larger
than the experimental mass as expected from the bound derived
in\cite{georgi}, and $\theta_P=-5.6^\circ$.
Note that the result differs from \cite{gl85}
because their mass matrix includes the term $k_2$ which we have
considered here to be of higher order.
In the mixed expansion the mixing angle is of order $O(1)$
rather than $O(m_q)$ as in the chiral expansion yet, numerically,
it comes out to be rather small at leading order. The ratio
${F_\pi}/{F_\eta}$ differs from one already at leading order and is
given by ${F_\pi}/{F_\eta}=1/\cos\theta_P$.

At order four, let us concentrate on the quantities which are
relevant in the discussion of the $\eta$ decay amplitude that is,
$\tan\theta_0$ and ${F_\pi}/{F_\eta}$. The former is given in closed
form by eq.(\ref{eq:mix3}) and the latter can be expressed as
\begin{equation}
{{F_\pi}\over{F_\eta}}={\cos(\theta_0-\theta_8)\over\lambda_8
\cos\theta_0}(1+z)
\end{equation}
As stated above, we must  expand the expressions
for $\tan\theta_0$ and ${F_\pi}/{F_\eta}$
linearly in terms of the $O(p^4)$ parameters
$y,z$ and the mass difference ${M_{\eta'}}^2 -
{\hbox{$M^2_{\eta'}$\kern-1.25em \raise 1.7ex\hbox{$\circ$}\kern.7em}}
$.
For this purpose, let us introduce the notations
\begin{equation}
\rho=\tan\theta_P,\ \qquad
\rho'={3{M_{\eta'}}^2-(2r_2+1){M_\pi}^2\over\sqrt2(r_2-1){M_\pi}^2}
\end{equation}
One finds for the ratio $F_\pi/F_\eta$
\begin{equation}
{F_\pi\over{F_\eta}}=
(1+\rho^2)^{1\over2}\left\{1-{2\over3}z(r_2-1)
+{r_2-1\over\sqrt2}\rho
(2y-{z\over3})-{1\over2}\rho^2{\rho\rho'-1\over\rho^2+1}\right\}
\label{eq:exp1}
\end{equation}
The first two terms in the parenthesis may be recognized
as those of the standard chiral expansion at $O(p^4)$ in
the leading $N_c\ $ approximation. Here, this formula is seen to
receive corrections.  The expansion of the mixing
angle, now:
\begin{equation}
-\tan\theta_0={1\over2}\rho(3-\rho\rho')+{r_2-1\over\sqrt2}(1-\rho^2)
(2y-{z\over3})
\label{eq:exp2}
\end{equation}

Let us now return to the question of the electromagnetic widths.
In the effective lagrangian language, the new feature compared
to sec.3 is that we have a piece containing $\phi_0$
\begin{equation}
{\cal L}_{\phi_0\to2\gamma}={8e^2\over F_0\sqrt{6(1+2k_1)}}
F_{\mu\nu}\tilde F^{\mu\nu}\phi_0
\left\{
{1\over32\pi^2 } +4k_3 +
{4(5+r_2)\over9} {M_\pi}^2 t_1\right\}
\label{eq:lagphi0}
\end{equation}
It can be seen that the amplitude for
${\eta'}\to2\gamma$ derived from (\ref{eq:lagphi0})
conforms to the general structure predicted in \cite{sv92}.

The lagrangian for $\phi_8$ is the same as before (\ref{eq:lageta}) with
the crucial difference that the $t_1'$ term is absent
(to be more precise, it no longer contains the $a'/{M_{\eta'}}^2$
piece in (\ref{eq:tprime}) and the rest being $O((N_c)^0)$
does not count at this order).  In order
to discuss the $\eta$ and ${\eta'}$ decays we expand $\phi_0$ and
$\phi_8$ using (\ref{eq:mixing}). In order to be consistent with the
$O(p^6)$ precision we use the mixing angles at $O(p^4)$ in the
$O(p^4)$ piece of the WZ lagrangian and the mixing angle $\theta_P$
in the piece which is already $O(p^6)$.

It is clear from (\ref{eq:lagphi0}) that we can make no definite prediction
for the decay of the ${\eta'}$. All we can do is use the experimental
information to constrain the combination of the
parameters $k_1$ and $k_3$ which appear
in (\ref{eq:lagphi0}). Once this is done,  we can make a prediction for the
$\eta$ decay. We can recast the eta decay amplitude in
a similar form as before (\ref{eq:aeta}):
\begin{equation}
A(\eta\to\gamma\gamma) ={\alpha\over\sqrt{3}\pi{F_\pi} }\left\{
{F_\pi\over F_\eta}
+{5-2r_2\over3\cos\theta_P}\,T_1\right\}
-\tan\theta_0\, A({\eta'}\to\gamma\gamma)
\label{eq:aetamix}
\end{equation}
Note that only one of the mixing angles shows up in this expression.
The expansions of $F_\pi/F_\eta$ and
$\tan\theta_0$ have been determined above (\ref{eq:exp1}),(\ref{eq:exp2}).
It is interesting to investigate the sensitivity of the
result upon small variations of the quark mass ratio $r$.
We will
consider the two cases $r=25.7$, as in \cite{gl85} and $r=23$ as
suggested from the corrections to Dashen's theorem. In the
first situation one obtains
\begin{equation}
{F_\pi\over F_\eta}=0.703,\qquad\theta_0=-20.4^\circ,\ \qquad
\Gamma(\eta\to\gamma\gamma)=439\ {\rm eV}\quad(r=25.7)
\end{equation}
This value of $r$ is the same as the one used in the chiral
expansion in sec.3 (standard case).
The value of the mixing angle obtained here is larger in
magnitude than the one obtained in sec.3.  One could a priori
expect that the the resulting $\eta$ width should be larger as
well but it turns out that this is not the case. Indeed, there
is an extra ingredient in the width which is ${F_\pi}/{F_\eta}$ and
using the mixed expansion rules we find a result which is smaller
than the chiral expansion result ${F_\pi}/{F_\eta}=0.77\pm0.03$. This
perhaps suggests that the uncertainty on this parameter is somewhat
underestimated. These two differences cancel out to some extent
and one ends up with practically the same result as before for
the $\eta$ width.
Now for the smaller value of $r$ we obtain
\begin{equation}
{F_\pi\over F_\eta}=0.710,\qquad\theta_0=-23.9^\circ,\ \qquad
\Gamma(\eta\to\gamma\gamma)=529\ {\rm eV}\quad(r=23.0)
\end{equation}
This illustrates that the uncertainty of the order of 100 eV
which was obtained in sec.3 can be traced, to a large extent,
to the uncertainty (of the order of 10\%) on the value of the
quark mass ratio $m_s/{\hat m}$ in the standard $\chi$pt.

\vspace{2mm}
\centerline{\bf 5. CONCLUSIONS}
\vspace{2mm}

In this paper we have discussed the chiral expansion
of the amplitudes for $\pi^0\to\gamma\gamma$ and
$\eta\to\gamma\gamma$ beyond leading order. The main new
contribution in this topic is the attempt made to evaluate
precisely, on the basis of sum-rules, the two parameters
$t_1$ and $t'_1$ which appear in ${\cal L}^{WZ}_{(6)}$.
The dominating
contribution, numerically, is $t'_1$ in good agreement with the fact
that it is $O(N_c^2)$ as it picks up a pole from the ${\eta'}$ meson.
We believe that our estimate of $t'_1$ should be on the same level
of reliability as the sum-rule for $L_7$. This sum-rule is, in fact,
one of the ingredients in the evaluation. The contribution from $t_1$
was found to amount for roughly 10\% of the width. Its evaluation
is certainly not as precise as for $t'_1$
but we argued that the sign and the
order of magnitude should be the correct ones. An interesting
peculiarity of this parameter is that it does not reflect the property
of a particular resonance but rather of the continuum, which is
matched to the QCD asymptotic behaviour.
It would be interesting to compare these results
with those which obtain in other approaches, in particular
improved variants of the NJL model which seem to perform well
for the standard $O(p^4)$ LE parameters\cite{njl}.
On the practical
side, we found that the $O(p^6)$ correction raises
the value of the width to a level which is compatible with the
photon-photon experimental results. The prediction
has a rather large uncertainty, of the order of 25\% which
is generated by the parameter $L_7$ and, to some extent,
by ${F_\pi}/{F_\eta}$. We have also investigated the relevance of
these two amplitudes in connection with the generalized $\chi$pt
which embodies values of $r$ much smaller than $r_2$. It turned
out that, even though individual factors are very different,
they combine to give a result which is also compatible with
experiment.

One of our aims was to test the convergence of chiral perturbation
theory in the anomalous sector. Since the lowest order result
is so small compared to experiment it was by no means obvious
how the $O(p^6)$ correction could
manage to bring the two in agreement.
This can be understood qualitatively if one assumes the simple
rule that a 1/$N_c\ $ correction has roughly the same magnitude as
a quark mass correction. The leading term in the
$O(p^6)$ correction
goes as $O(N_c^2)\times O(m_q)$ so, according to this rule, it is
natural to expect that it could be of a similar size as the
leading chiral order contribution which goes as $O(N_c)\times O(1)$.
Terms of still higher order, on the other hand, should be
much smaller. On the basis of this counting scheme it is possible
to include the $\eta'$ into the effective lagrangian in a systematic
way. We have discussed the question of the $\eta$ decay from this
mixed expansion point of view. In a sense, this proves slightly
disappointing because, even in this scheme, the essential contribution
to the mixing angle comes at next to leading order. The mixing angle
turns out to be slightly larger than in the chiral expansion but the
predictions for the width are nearly identical.
There remains to explore whether this
kind of expansion provides some non-trivial constraints in the $\eta'$
sector.

\vspace{2mm}
\noindent{\bf Acknowledgements:}
Jan Stern is thanked for numerous suggestions, explanations
and discussions, and Marc Knecht  and Hagop Sazdjian for
discussions.

\end{document}